\documentclass{ws-ijmpb}
\usepackage{bm}

\begin{document}

\markboth{Pei-Ling Zhou,  Zi-Nan Tang,  Tao Zhou,  Jing-Ting Wang,  and  Chun-Xia Yang}
{Mathew Effect in Artificial Stock Market}

\catchline{}{}{}{}{}

\title
{Mathew Effect in Artificial Stock Market}

\author
{Pei-Ling Zhou,  Zi-Nan Tang,  Tao Zhou,  Jing-Ting Wang,  and Chun-Xia Yang}

\address
{Department of Electronic Science and Technology,\\
University of Science and Technology of China,\\
Heifei Anhui, 230026, PR China\\}

\maketitle

\begin{abstract}
In this article, we established a stock market model based on agents' investing mentality. 
The agents decide whether to purchase the shares at the probability, according to their anticipation of
 the market's behaviors. The expectation of the amount of shares they want to buy is directly proportional
 to the value of asset they hold. The agents sell their shares because of the gaining-profit psychology, 
stopping-loss psychology, or dissatisfaction with the long-time congealing of the assets. We studied how the
 distribution of agent's assets varies along with systemic evolution. The experiments show us obvious Mathew
 effect \footnote{¡°Mathew Effect¡± is a social psychology effect firstly put forward by American science history researcher
 {\it R. Merton} in 1973, which means ¡°the more renowned scientists are, the more honor they receive, while the achievement made
 by those unknown scientists are not admitted.¡± Thenceforward, the ¡°Mathew Effect¡± conception was introduced to economics, 
which was used to describe predominance accumulating phenomena in social wealth distribution and economic development.}
 on asset distribution in the artificial stock market, and we have found that the Mathew effect on asset 
distribution was more and more salient along with the increasing of system running time, stock market size and agents' activity extent.
\end{abstract}


\section*{}
\noindent Financial Market dynamics emerging from a large number of interacting agents have raised considerable interest\cite{1}. Recently,
 fruitful attempts have been made to understand such emergent phenomena in systems composed of many interacting agents, 
and some simple models 
which can, in principle, capture essential features of real market are established\cite{2,3,4}. Two notable models are SFI's artificial 
stock market\cite{5} and Lux-Model\cite{6,7}. The two models reproduce several nontrivial properties observed in real markets, 
therefore are quite successful. But from the theoretical point of view, their common drawback is highly complex with too 
many free parameters and considerable tuning. Thus some econophysicists attempt to capture some key features of a genetic
 market mechanism using drastic simplification model, such as minority game\cite{8,9,10}, 
El Farol's Bar problem\cite{11}, spin model\cite{12,13,14} and so on. 
 In this article, a very simple stock market model based on agents' investing mentality is established. 
The obvious Mathew effect on asset distribution is found, and is testified to become more salient along with the increasing 
of system running time, stock market size and agents' activity extent.

In this model, agents submit their buying and selling applications according to their own trading strategies. 
Each application includes application trading price and amount. All of agents' applications are separately 
classified into the buyer-sequence and seller-sequence. In the buyer-sequence, application trading prices
 are arranged in non-increasing order£¬while in the seller-sequence they are arranged in non-decreasing order. 
The former application in the sequence will be carried out preferentially. The trading price is the average of the 
application buying price and application selling price. 

In each time step, agents make the buying application at the probability of $b$, where $b \in  (0,1)$
 is called purchase factor, and they will sell the shares, if one of the following three conditions is satisfied: 

\begin{romanlist}[(ii)]
\item When the stock price $P$ is higher than share-holding price $P'$, and agents have gained more income from the stock than their expectation,
 they will consider this invest as a success, and sell the stock. The selling condition is $P'>P\times (1+e)$;
\item When the stock price $P$ is lower than share-holding price $P'$, and agents are unbearable because of the loss they have suffered, 
and then they will
 also sell the stock in order to avoid a larger loss. The selling condition is $P'<P\times (1-e)$;
\item When agents' share-holding time $d'$ is longer than maximal share-holding time $d$ they can bare, 
 they will sell the stock and plough the fund into the next invest circle
in order to avoid leaving fund unused. The selling condition is $d'>d$.  
\end{romanlist}

Where $e\in (0,1)$ is called price-accept factor, $d>0$ is called the maximal share-holding time.

The buyer will choose $1/3$, $2/3$ or entire of the assets to buy, and the seller will choose $1/3$, $2/3$ or 
entire of shares to sell, the choice is made up randomly. The application buying price is a little higher
 than the current stock price, while the application selling price is a little lower. 

There are four parameters in this model: the number of agents $N$ stands for market size, 
and agent attribute $b$, $e$ and $d$ describe the investing strategy and risk-fancy. 
Purchase factor stands for the desire agents want to buy. When $b=1$, agent has 
the highest purchase desire and will submit buying application in each time step; 
and $b=0$ means agent have exit the stock market. When the average of all the purchase 
factors is large enough, stock market become into long position market, otherwise nominal market. 
When price-accept factor is somewhat large, agent acts as a risk-partiality investor, 
who has comparative speculation. Contrarily, agent acts as a risk-elusion investor, who
 is sensitive at the fluctuation of stock price, little profit or loss will urge her to trade. When 
the average of all the price-accept factors is high enough, there are so many speculators in the market, 
who make the market be in high-risk state. Contrarily, agents invest discreetly and the market is in low-risk state. 
The value of $d$ marks off the long-term investors and short-term investors, and its average value describes the speed 
of fund-turnover in the market.

Agents' assets are an important index to weigh agents' status in the stock market. 
At the beginning, let the quantity of each agent's fund and shares be the same, as the market running, polarization 
phenomena of agents' assets will be observed. Some agents benefit in trades, whose wealth increased, comparatively, 
other agents' wealth decreased because of wrong strategy. Fig. 1 shows that the polarization becomes more and more 
obvious as the market running. The higher asset-value will endow winners larger chance to benefit, which help them keep
 the advantage. Meanwhile, losers' assets are more possible to decrease than others, which keep them at the bottom for a long time.
 The disparity between the poor and rich becomes larger and larger, the distribution of agents' assets shows Mathew
 effect, which appears the rich richer, the poor poorer. 

\vspace{2mm}

\centerline{\scalebox {0.7} [0.3] {\includegraphics {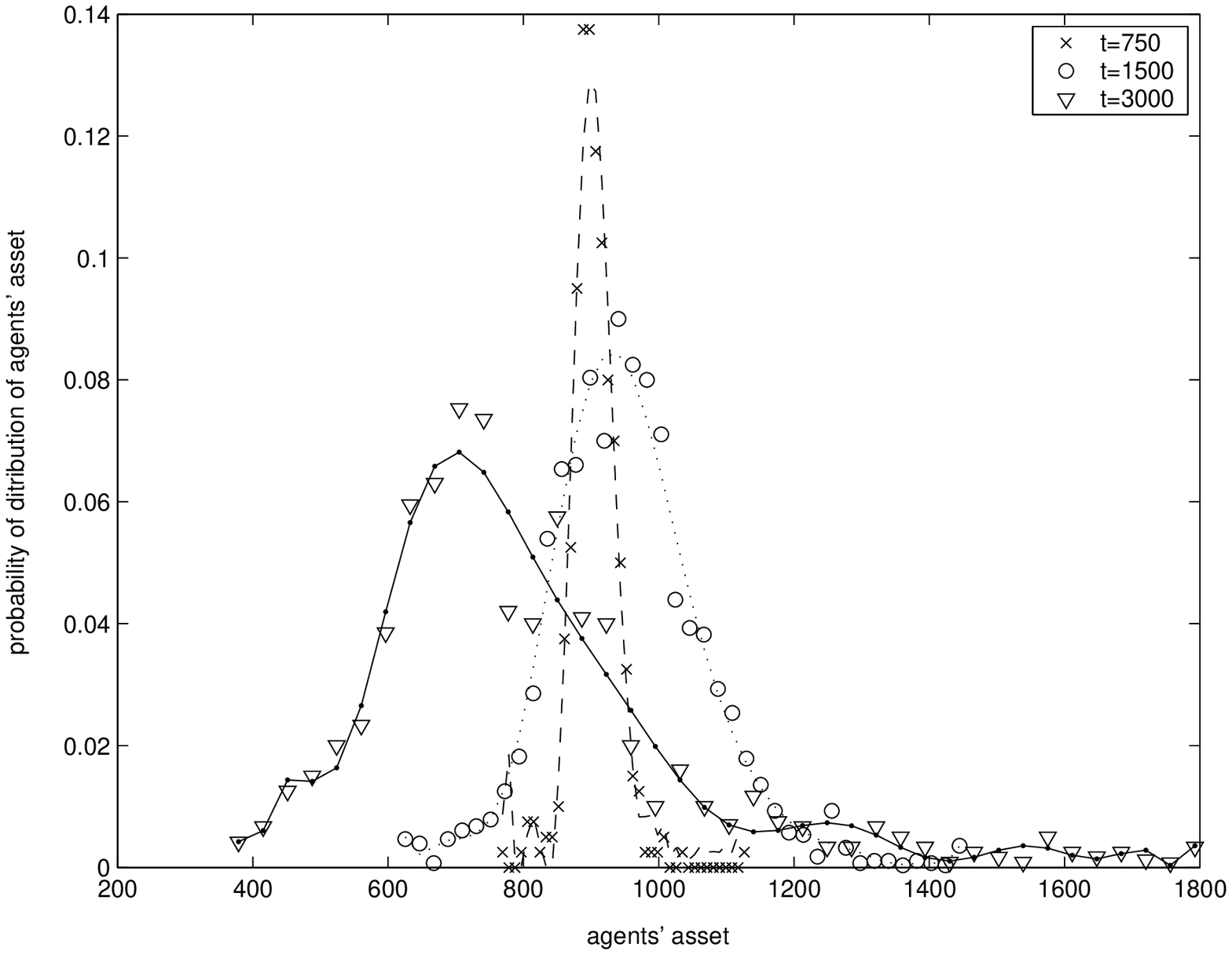}}}

\centerline{{\bf Fig. 1.} Distribution of agents' asset. The market-size is 400, each agent has}
\centerline{asset as 950.0 in the beginning. When $t=750$, the maximal asset is 1125.79,}
\centerline{and the minimal asset is 793.41, as $t=1500$, they become 1452.12 and 617.95,}
\centerline{and after 3000 time steps, they polarize to be 1781.33 and 371.40.}

\vspace{3mm}

In order to depict the Mathew effect in artificial stock market quantificationally, 
we studied the range of agents' asset in the stock market by experiments.
 Let $m_i$, $q_i$ be the money and stock amount of agent $i$, $P_t$ be the stock price in time step $t$, then the range 
$R_t$ of agents' asset in time step $t$ are defined as:

\[
R_t=\max\{m_i+q_i \times P_t | 1 \leq i \leq N\}-\min\{m_i+q_i \times P_t | 1 \leq i \leq N\}
\]

\vspace{1mm}

\centerline{\scalebox {0.7} [0.3] {\includegraphics {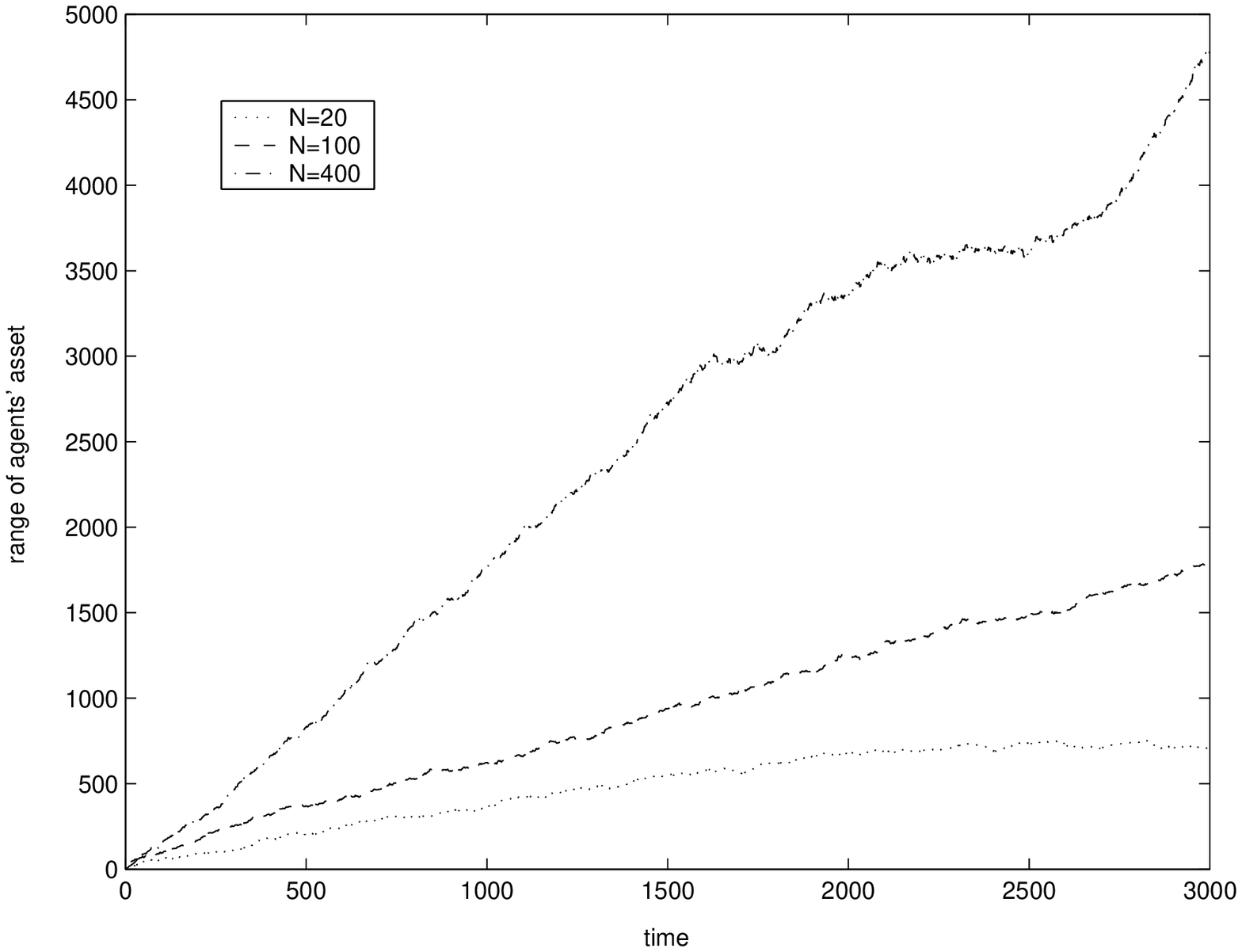}}}
\centerline{{\bf Fig. 2.} $N=20$, 100, 400: the range of agents' assets as a function of time.}

\vspace{3mm}

Fig. 2 shows how the range of agents' assets vary along with the time 
with each agent's asset is 950.0 at the beginning. In the figure, the disparity between the poor and rich becomes
 larger and larger as time goes by. For a stock market of 400 agents, after the system ran 3000 time steps,
 the range of asset has come to 4831.77. In fact, after 2307 days, the agent who holds the least asset hasn't 
had enough money to make any trade, which means she has been washed out of the market. It can be supposed 
that Mathew effect has the current of continuing accretion, if only there are enough agents able to trade in the market.

\vspace{4mm}

\centerline{\scalebox {0.7} [0.3] {\includegraphics {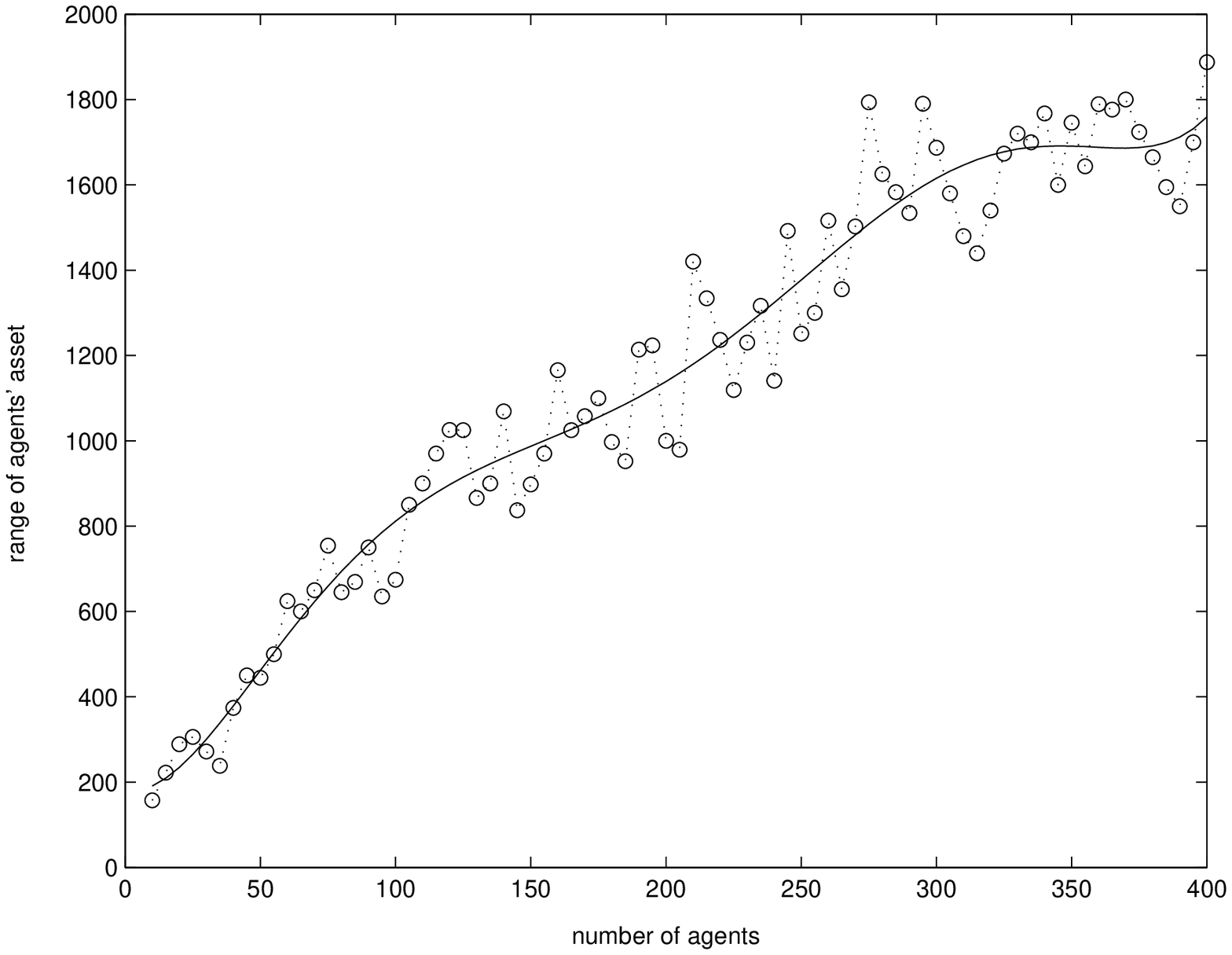}}}
\centerline{{\bf Fig. 3.} $t=1000$: the range of agents' assets as a function of market size.}

\vspace{3mm}

We can observe from fig. 2 that, the larger the market-size is, the more obvious the Mathew effect is. 
Fig. 3 describes how the market-size influences Mathew effect in stock market. On the condition that agents have 
the same initial asset and attribute, according to the range of agents' assets after the system ran 1000 time steps, 
the positive correlation between Mathew effect and the market-size can be found. And along with the increase of market size, 
Mathew effect becomes more and more obvious in this stock market.

\vspace{4mm}

\centerline{\scalebox {0.7} [0.3] {\includegraphics {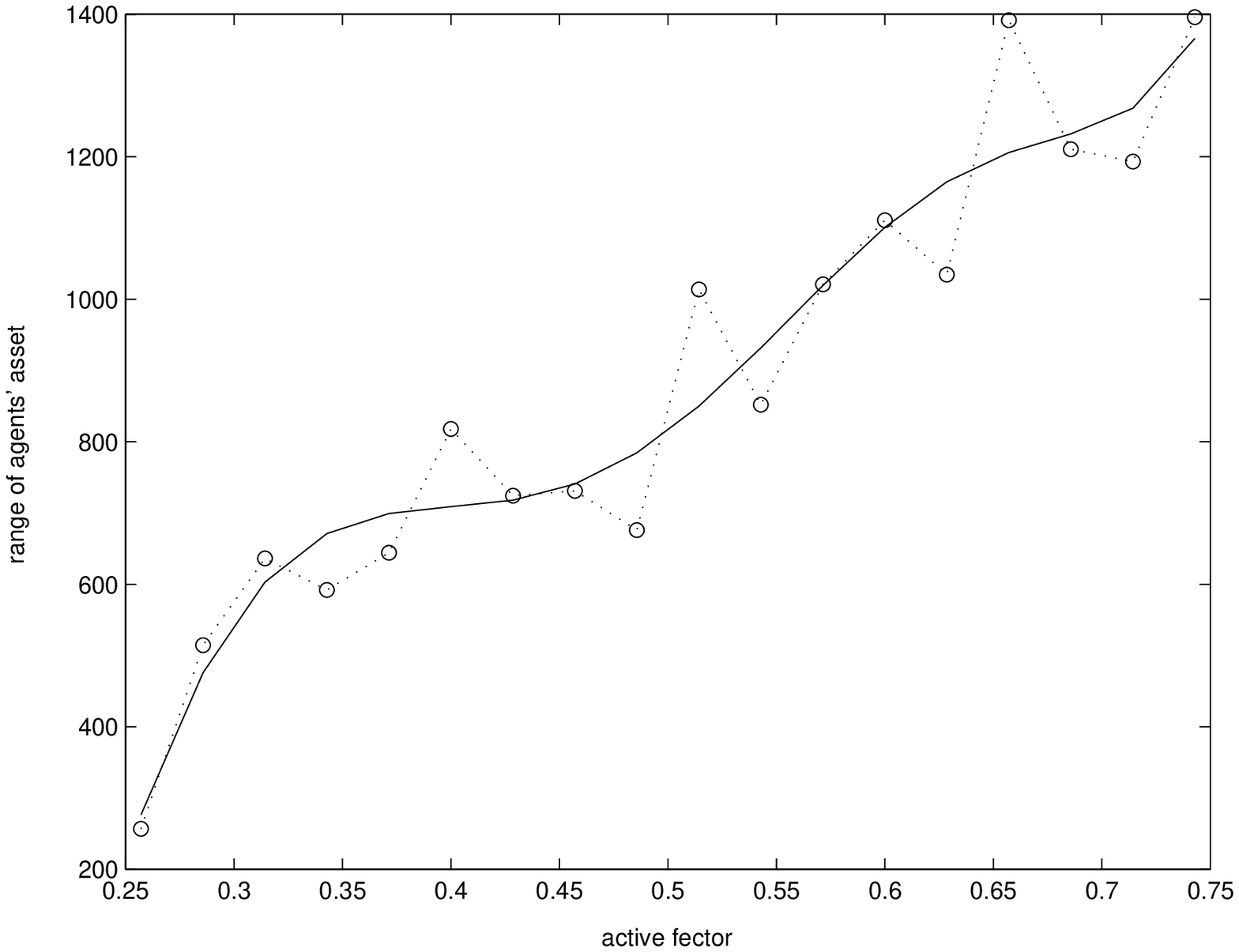}}}
\centerline{{\bf Fig. 4.}  $N=200$, $t=1000$: the range of agents' asset as a function of agents' active factor.}

\vspace{3mm}

Another problem worthwhile to discuss is how agents' investing strategies influence the Mathew effect in stock market.
 Considering two extreme conditions, when the purchase factor is very small while price-accept 
factor and the maximal share-holding time is very large, agents will apply to buy or to sell shares at very small probability. 
On the contra, agents will put forward application frequently, which has weighty influence on the behavior 
of the stock market system. If $t_i$, $e_i$, $d_i$ stand for the purchase factor, price-accept factor and the maximal share-holding 
time of agent $i$, the active factor $a(a\in (0,1))$ of agent $i$ is defined as: 

\[
a_i=\frac{1}{3}[ b_i + (1-e_i) + \frac{2}{\pi} \arctan d_i]
\]  

Fig. 4 describes how the range of agents' asset varies along with the average of agents' active factor. 
We can see that Mathew effect in the stock market with more active agents is more obvious, 
which means large trade-amount and frequent trade will widen the gap between the poor and the rich.

In the resent years, Mathew effect has attracted widespread attention of economists. 
In this article, we have established a simple stock market model which shows us obvious Mathew 
effect on asset distribution. And some interesting phenomena similar to the real stock market are discovered by
 experiments. It is significant to carry out the research on how to explain the cause inducing Mathew effect and to
 gauge the magnitude that Mathew effect brings to us.

\section*{Acknowledgements}

\noindent Supported by the National Natural Science Foundation of China under Grant No. 70171053.


\begin{thebibliography}{0}

\bibitem{1}
G. J. Stigler, {\it J. Business} {\bf 37}, 117 (1964).

\bibitem{2} 
M. Levy, H. levy, and S. Solomon, {\it Econ. Lett.} {\bf 45}, 103 (1994).

\bibitem{3} 
M. Youssefmir, B. Huberman, {\it J. Econ. Behav. Organization} {\bf 32}, 101 (1997).

\bibitem{4}
R. Cont, J. -P. Bouchaud, {\it Macroecon. Dynamics} {\bf 4}, 170 (2000).

\bibitem{5}
W. B. Arthur, J. Holland, B. LeBaron, et al. 
{\it Asset pricing under endogenous expectations in an artificial stock market.}
 In W. B. Arthur, S. Durlauf, D. Lane ed. 
{\it The Economy as an Evolving Complex System I$\!\!$I}
 (Boston: Addison-Wesley, 1997, pp. 15-44).

\bibitem{6}
T. Lux, M. Marchesi, {\it Nature} {\bf 397}, 498 (1999).

\bibitem{7}
T. Lux, M. Marchesi, {\it Int. J. Theor. Appl. Finance} {\bf 3}, 675 (2000).

\bibitem{8}
D. Challet, Y. -C. Zhang, {\it Physica} {\bf A 246}, 407(1997). 

\bibitem{9}
R. Savit, R. Manuca, and R. Riolo, {\it Phys. Rev. Lett.} {\bf 82}, 2203(1999).

\bibitem{10}
D. Challte, A. Chessa, M. Marsili, and Y. -C. Zhang, {\it Quantitative Finance} {\bf 1}, 168(2001).

\bibitem{11}
W. B. Arthur, {\it Am. Econ. Rev.} {\bf 84}, 406(1994).

\bibitem{12}
G. Iori, {\it Int. J. Mod. Phys.} {\bf C 10}, 1149(1999).

\bibitem{13}
D. Chowdhury, D. Stauffer, {\it Eur. Phys. J.} {\bf B 8}, 477(1999).

\bibitem{14}
Stefan Bornholdt, {\it Int. J. Mod. Phys.} {\bf C 12}, 667(2001).

\end{thebibliography}
\end{document}